\documentclass[aps,prb,twocolumn,superscriptaddress,showpacs]{revtex4-1}

\usepackage{graphicx}
\usepackage{calc}
\usepackage{bm}
\usepackage{color}
\usepackage{textcomp}

\begin{document}

\title{Fractional ac Josephson effect as evidence of topological hinge states in a Dirac semimetal NiTe$_2$}

\author{D.Yu.~Kazmin}
\author{V.D.~Esin}
\author{A.V.~Timonina}
\author{N.N.~Kolesnikov}
\author{E.V.~Deviatov}

\affiliation{Institute of Solid State Physics of the Russian Academy of Sciences, Chernogolovka, Moscow District, 2 Academician Ossipyan str., 142432 Russia}

\date{\today}

\begin{abstract}
  We  experimentally investigate Josephson current between two 5~$\mu$m spaced superconducting indium leads, coupled to a NiTe$_2$ single crystal flake, which is  a type-II Dirac semimetal.  Under microwave irradiation, we demonstrate a.c. Josephson effect at millikelvin temperatures as a number of Shapiro steps. In addition to the integer ($n=1,2,3,4...$) steps, we observe fractional ones at half-integer values $n=1/2,3/2,5/2$ and 7/2, which corresponds to $\pi$ periodicity of current-phase relationship. In contrast to previous investigations, we do not  observe $4\pi$ periodicity (disappearance of the odd $n=1,3,5...$ Shapiro steps), while the latter is usually considered as a fingerprint of helical surface states in Dirac semimetals and topological insulators.  We argue, that our experiment confirms Josephson current through the topological hinge states in  NiTe$_2$: since one can exclude bulk supercurrent in 5~$\mu$m long Josephson junctions, interference of the hinge modes is responsible for the $\pi$ periodicity, while stable odd Shapiro steps reflect chiral character of the topological  hinge states. 
\end{abstract}


\maketitle

\section{Introduction}

Like other topological materials~\cite{Volkov-Pankratov,hasan,zhang,das,chiu}, topological semimetals acquire topologically protected  surface states due to the bulk-boundary correspondence~\cite{armitage}. In contrast to topological insulators, Dirac semimetals are characterized by gapless bulk spectrum with band touching in some distinct Dirac points. In Weyl semimetals every touching point splits into two Weyl nodes with opposite chiralities. Fermi arc surface states  are  connecting projections of these nodes on the surface Brillouin zone, so the topological surface states are chiral for Weyl materials~\cite{armitage}. For Dirac semimetals, the surface states are helical, similarly to topological insulators~\cite{Volkov-Pankratov}. 

The main problem of transport investigations is to reveal the surface states contribution in topological semimetals with gapless bulk spectrum~\cite{armitage}. In proximity with a superconductor, topological surface (or edge) states are able to carry supercurrents over extremely large distances~\cite{topojj1,inwte1,inwte2,topojj3,incosns,topojj5,rfli}, while the coherence length is much smaller for the bulk carriers. Also, nonuniform supercurrent distribution is  reflected in dc or ac Josephson effect.  For example, it may lead to the superconducting quantum interference device (squid)-like critical current suppression pattern~\cite{yakoby,kowen,inwte1} and/or to the fractional a.c. Josephson effect~\cite{inwte2,rfpan,rfsnyder} with half-integer Shapiro steps. 

For the typical Dirac semimetal Cd$_3$As$_2$, observation  of $\pi$ and $4\pi$ periodic current-phase relationship has been reported  in Al-Cd$_3$As$_2$-Al and Nb-Cd$_3$As$_2$-Nb  junctions~\cite{rfli,rfpan}.  For the short 100~nm junctions, the fractional a.c. Josephson effect ($\pi$ periodicity) is connected with interference between the bulk and surface supercurrent contributions~\cite{rfpan}, while the disappearance of $n=1$ Shapiro step ($4\pi$ periodicity)  reflects  the helical nature of topological surface states in Dirac semimetals also for 1~$\mu$m long junctions~\cite{rfli}.
 
Recently it has been understood, that besides the well-known three-dimensional bulk Dirac states and the two-dimensional Fermi-arc surface states, there should be one-dimensional hinge states~\cite{hinge_review} at the intersections between surfaces of Dirac semimetals~\cite{hinge_theory,hingeCdAs_theory}.  Dirac semimetals exhibit hinge states as universal, direct consequences of their bulk three-dimensional Dirac points, see Ref.~\cite{hinge_review} for details. These hinge states represent a new kind of Chern-type insulator edge states, so they are chiral even for Dirac semimetals~\cite{hinge_review,hinge_theory}.  The idea of the experiment~\cite{hingeCdAs_experiment} is  to distinguish between different types of current-carrying states since  the coherence length should be longer in the one-dimensional hinge channel than that in the surface and bulk ones. As a result,  standard Fraunhofer interference pattern in a short  Josephson junction is changed to the squid-like one for the 1~$\mu$m long Cd$_3$As$_2$ based junctions~\cite{hingeCdAs_experiment}, because the supercurrent is dominated by several hinge channels in the latter case.

Due to the topological origin, the effect should also be  independent on the particular material. 
NiTe$_2$ is a recently discovered type-II  Dirac semimetal belonging to the family of transition metal dichalcogenides. Nontrivial topology of NiTe$_2$ single crystals has been confirmed by spin-resolved ARPES~\cite{PhysRevB.100.195134, Mukherjee2020}. The bulk coherence length $\xi$  is smaller in NiTe$_2$ in comparison with Cd$_3$As$_2$ due to the smaller mean free path $l_e$, which should further suppress the bulk supercurrent. The contribution from the topological surface states reveals itself as the Josephson diode effect in parallel magnetic field~\cite{JDE,aunite}. 

Similarly to Cd$_3$As$_2$ Dirac material, hinge states have been theoretically predicted~\cite{NiTe2_hinge} in NiTe$_2$. Hinge supercurrent has been demonstrated in the submicron-size NiTe$_2$ based Josephson junction as a (squid)-like critical current suppression pattern due to the magnetic field suppression of the bulk supercurrent~\cite{NiTe2_hinge}. Thus, it is reasonable to study a.c. Josephson effect in long NiTe$_2$ junctions to confirm topological (chiral) nature of the predicted hinge states in Dirac semimetals.

Here, we experimentally investigate Josephson current between two 5~$\mu$m spaced superconducting indium leads, coupled to a NiTe$_2$ single crystal flake, which is  a type-II Dirac semimetal.  Under microwave irradiation, we demonstrate a.c. Josephson effect at millikelvin temperatures as a number of Shapiro steps. In addition to the integer ($n=1,2,3,4...$) steps, we observe fractional ones at half-integer values $n=1/2,3/2,5/2$ and 7/2, which corresponds to $\pi$ periodicity of current-phase relationship. In contrast to previous investigations, we do not  observe $4\pi$ periodicity (disappearance of the odd $n=1,3,5...$ Shapiro steps).

\section{Samples and technique}

\begin{figure}
\includegraphics[width=\columnwidth]{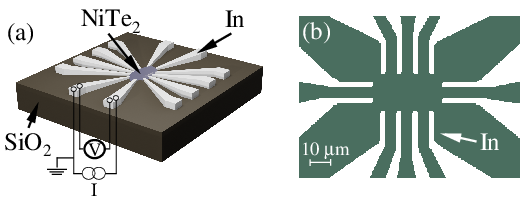}
\caption{(Color online) (a) A sketch of a sample with In leads and external electrical connections. A thick (above 0.5~$\mu$m) NiTe$_2$ mechanically exfoliated flake is placed on the  pre-defined In leads pattern to form 5~$\mu$m separated In-NiTe$_2$-In junctions. Electron transport is investigated between two neighbor superconducting indium leads in a four-point technique,  all  the wire resistances are excluded. (b) An optical image of the indium leads pattern before transferring the NiTe$_2$ single crystal flake.
  }
\label{fig1}
\end{figure}

NiTe$_2$ was synthesized from elements, which were taken in the form of foil (Ni) and pellets (Te). The mixture was heated in an evacuated silica ampule up to 815$^{\circ}$ C with the rate of 20 deg/h, the ampule was kept at this temperature for 48 h. The crystal was grown in the same ampule by the gradient freezing technique with the cooling rate of 10 deg/h. As a result, we  obtain  80~mm long and 5~mm thick NiTe$_2$ single crystal,  with (0001) cleavage plane. 

The powder X-ray diffraction analysis (Cu K$\alpha$1 radiation, $\lambda$ = 1.540598 \AA) confirms single-phase NiTe$_2$ with P-3m1 (164) space group (a = b = 3.8791 \AA, c = 5.3005 \AA), see Fig.~\ref{fig1}(a).  The known structure model  is also refined with single crystal X-ray diffraction measurements (Oxford diffraction Gemini-A, Mo K$\alpha$). Nearly stoichiometric ratio Ni$_{1-x}$Te$_2$ ($x$ $<$ 0.06) is verified by the energy-dispersive X-ray spectroscopy. 

The quality of our NiTe$_2$ material was also tested in standard four-point magnetoresistance measurements, see Ref.~\onlinecite{aunite} for details. In particular, nonsaturating longitudinal magnetoresistance~\cite{Xu2018,PhysRevB.99.155119} was confirmed for our NiTe$_2$ samples in normal magnetic field~\cite{aunite}. It is important, that four-point resistance is finite (0.1~$\Omega$) between two 5~$\mu$m spaced Au leads in zero magnetic field, so there is no bulk superconductivity~\cite{Feng2021} for  NiTe$_2$ single crystal flakes at ambient pressure even at millikelvin temperatures~\cite{aunite}. 

Fig.~\ref{fig1} (a) shows a sketch of a sample. Despite NiTe$_2$ can be thinned down to  two-dimensional monolayers, topological semimetals are essentially three-dimensional objects~\cite{armitage}. Thus, we have to select  relatively thick (above 0.5~$\mu$m) NiTe$_2$ single crystal flakes, which also ensures sample homogeneity. 

Thick flakes require special contact preparation technique: the In leads pattern is firstly formed on a standard oxidized silicon substrate by lift-off, as depicted in Fig.~\ref{fig1} (b). The 100~nm thick In leads are separated by 5~$\mu$m intervals, which defines the experimental geometry. As a second step, the fresh mechanically exfoliated NiTe$_2$ flake is transferred to the In leads pattern and is shortly pressed  to the leads by another oxidized silicon substrate, the latter is removed afterward. The substrates are kept  strictly parallel by external metallic frame to avoid sliding of the NiTe$_2$ crystal, which is verified in optical microscope. As a result,  planar In-NiTe$_2$ junctions  are formed at the bottom surface of the NiTe$_2$ single crystal flake in Fig.~\ref{fig1} (a), being separated by 5~$\mu$m intervals, as depicted in Fig.~\ref{fig1} (b). As an additional advantage, the In-NiTe$_2$ junctions and the surface between them are protected from any contamination by SiO$_2$ substrate, since they are placed at the bottom side of a thick NiTe$_2$ flake in Fig.~\ref{fig1} (a). 

This procedure provides transparent In-NiTe$_2$ junctions, stable in different cooling cycles, which has been verified  before for a wide range of materials~\cite{aunite,inwte1,inwte2,incosns,infgt,ingete}.  Thus, they are suitable to form long In-NiTe$_2$-In  SNS structures. The mean free path $l_e$ can be estimated as $l_e\approx$4~$\mu$m in NiTe$_2$ from the four-point resistance~\cite{aunite}, so it is smaller than the  $L=5 \mu$m intervals between indium leads in Fig.~\ref{fig1}. $L$ should be compared~\cite{kulik-long,dubos} with the coherence length of the diffusive SNS junction $\xi=(l_e \times \hbar v_F^N/\pi\Delta_{in})^{1/2}\approx 300$~nm, where Fermi velocity is  $v_F^N\approx 10^7 \frac{cm}{s}$, and $\Delta_{In}=0.5$~meV is the indium superconducting gap~\cite{indium}. Due to the obvious relation $L/\xi > 10$, one can not expect bulk Josephson current in our $L=5 \mu$m long In-NiTe$_2$-In  SNS structures. This estimation well corresponds to the results of Ref.~\onlinecite{aunite}, where there was no bulk contribution to the supercurrent on the pristine NiTe$_2$ surface between two 1~$\mu$m spaced superconducting leads.

We study electron transport between two superconducting indium leads in a four-point technique. An example of electrical connections is shown in Fig.~\ref{fig1} (a): one In electrode  is grounded, the current $I$ is fed through the neighboring one; a voltage drop $V$ is measured between these two indium electrodes by independent wires. In this connection scheme, all  the wire resistances are excluded, which is necessary for low-impedance  In-NiTe$_2$-In junctions ( 0.25--5~Ohm normal resistance in the present experiment). 

The indium leads are superconducting below the  critical temperature~\cite{indium} $T_c\approx 3.4~K$. However, we observe Josephson current only below 1~K, so the measurements are performed in a dilution refrigerator with 30~mK base temperature.  For a.c.  Josephson effect investigations, the sample is illuminated by microwave (rf) radiation through an open coaxial line. Due to specifics of the dilution frige, we have to restrict rf radiation by 0.5~GHz frequency and 7~dBm power, the bath temperature is always below 60~mK in this case.

\begin{figure}
\includegraphics[width=\columnwidth]{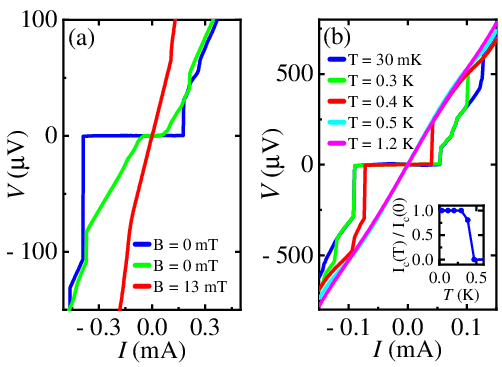}
\caption{(Color online) Standard Josephson $I-V$ characteristics for three different In-NiTe$_2$-In samples with different critical currents and normal resistances (blue and green curves in (a), blue curve in (b)), obtained at 30~mK temperature in zero magnetic field. The current sweep direction is from the positive to negative currents, which is the origing of $I-V$ asymmetry. 
Suppression of the zero-resistance state  is shown in (a) for $B=13$~mT  normal to the plane magnetic field (red curve) at 30~mK temperature and in (b) for different temperatures in zero field. Inset shows the normalized critical current temperature dependence $I_c(T)/I_c(0)$.   
 }
\label{fig2}
\end{figure}

\section{Experimental results}

To obtain $I-V$ characteristics,  we sweep the dc current $I$ and measure the voltage drop $V$, see  $I-V$ curves in Fig.~\ref{fig2} for three different samples. The normal In-NiTe$_2$-In  junction resistance varies from $\approx 0.2$~$\Omega$ in (a)  to 5~$\Omega$ in (b), due to the different overlap between the NiTe$_2$ flake and In leads for different samples.  

In zero magnetic field and at low 30~mK temperature, Fig.~\ref{fig2} shows standard Josephson behavior, despite of $L>>\xi$ for the present In-NiTe$_2$-In junctions: (i) by the four-point connection scheme we directly demonstrate zero resistance  region at low  currents. (ii) The non-zero resistance appears as sharp jumps  at current values $ I_c\approx 0.05 - 0.4$~mA for different samples. The current sweep direction is from the positive to negative currents, so the cricical Josephson current is characterized by left (negative) $I-V$ branch. The obtained $I_c$ values are much smaller than the critical current for the indium leads, which can be estimated as  $\approx 30$~mA  for the leads' dimensions and the known~\cite{in-current} indium critical current density $j\approx 3\times 10^6$A/cm$^2$. (iii) $I-V$ curve can be switched to standard Ohmic behavior, if  the supercurrent is suppressed by magnetic field (13~mT) or temperature (above 0.5~K), see (a) and (b), respectively. 

Inset to  Fig.~\ref{fig2} (b) shows the normalized critical current temperature dependence $I_c(T)$, which is unusual for long $L>\xi$ diffusive $L>>l_e$ SNS junctions~\cite{kulik-long,dubos}. On the other hand, one can not expect bulk Josephson current for  $L/\xi > 10$ for our In-NiTe$_2$-In  structures, while the topological surface states  carry the Josephson current in Refs.~\onlinecite{JDE,aunite,NiTe2_hinge}  probably due to the backscattering suppression.

\begin{figure}
\includegraphics[width=\columnwidth]{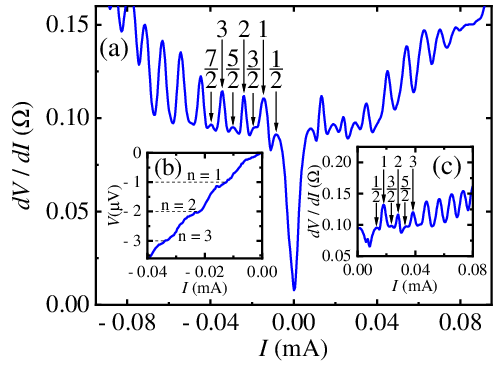}
\caption{(Color online) Differential $dV/dI(I)$ characteristics of the In-NiTe$_2$-In  junctions under rf (0.5~GHz, 7~bBm) irradiation through an open coaxial line. The data are obtained  in zero magnetic field, the sample temperature is always below 60~mK. Shapiro steps with integer numbers $n=1,2,3,4,...$ can be seen as sharp equidistant $dV/dI$ peaks. We check, that the peaks well correspond to the centers of the steps in usual dc $I-V$ curves (e.g. in the left inset, where Shapiro steps at $V=n hf/2e$ levels are depicted by dashed lines).
 In addition to the integer steps, there are fractional ones at  $n=1/2,3/2,5/2$ and 7/2 as small $dV/dI$ peaks, some $dV/dI(I)$ asymmetry is connected with the current sweep direction. Right inset shows qualitatively similar $dV/dI(I)$ behavior with half-integer Shapiro steps for another sample, 
 so the main experimental finding is the observation of fractional a.c. Josephson effect with $\pi$ periodicity  of the current-phase relationship.  
 }
\label{fig3}
\end{figure}

The main experimental finding is the observation of fractional a.c. Josephson effect with $\pi$ periodicity  of the current-phase relationship. 

Differential $dV/dI(I)$ characteristics under rf irradiation are shown in in Fig.~\ref{fig3}. Dilution fridge  restricts radiation frequency (0.5~GHz) and power (7~dBm) to avoid sample overheating through the coaxial lines. The power at the sample is unknown, but the base temperature is always below 60~mK under irradiation.  In these conditions, $dV/dI(I)$ curves allows to increase resolution of Shapiro steps in Fig.~\ref{fig3}: the dc current $I$ is additionally modulated by a low (100~nA) ac component, the ac ($\sim dV/dI$) voltage is measured  by a lock-in amplifier. The signal is confirmed to be independent of the modulation frequency within 100 Hz -- 10kHz range, which is defined by the applied filters. 

Shapiro steps with integer numbers $n=1,2,3,4,...$ can be seen as sharp equidistant $dV/dI$ peaks, see the main field of Fig.~\ref{fig3}. We check, that they are the peaks which well correspond to the centers of the steps in usual dc $I-V$ curves, see the left inset to Fig.~\ref{fig3}. Thus, we should concentrate on the peaks (not dips) while considering differential $dV/dI(I)$ characteristics, probably due to the imperfect steps' shape  at low frequency and power. Shapiro steps  are placed at  $V=n hf/2e$ in the inset to Fig.~\ref{fig3}, as it should be expected for typical SNS junctions with trivial $2\pi$ periodicity in current-phase relationship $I_J\sim sin(\phi)$. 

In addition to the integer steps, we observe small but clearly visible fractional ones at  $n=1/2,3/2,5/2$ and 7/2 as small $dV/dI$ peaks. The peaks appear at low currents $I$, they can be seen both for positive and negative currents. Some $dV/dI(I)$ asymmetry is connected with the current sweep direction from the positive to negative current values, so the right $dV/dI(I)$ branch is obtained while sweeping from the resistive sample state. Thus, we demonstrate $\pi$ periodicity  of the current-phase relationship for the fractional a.c. Josephson effect in long In-NiTe$_2$-In junctions. 

This behavior is well-reproducible for different samples, e.g. $dV/dI(I)$ curve under rf irradiation is presented in the right inset to Fig.~\ref{fig3} with integer and half-integer Shapiro steps. It is important, that this reproducibility is  inconsistent with sample fabrication defects, e.g. parasite shorting of In leads. As additional arguments against fabrication defects, the thickness of the indium film is chosen to be much smaller than the leads separation (100~nm$<< 5\mu$m) to avoid parasite shortings. The zero-field critical current values for the present In-NiTe$_2$-In  junctions well correspond to ones in Ref.~\cite{aunite}, despite different materials of the superconducting leads.  Josephson effect is fully suppressed  above 0.5~K and at 13~mT magnetic field in Fig.~\ref{fig2}, which is also far below the values for pure indium~\cite{indium}. 
Thus, we should conclude that long  In-NiTe$_2$-In  junctions show half-integer steps ($\pi$ periodicity  of the current-phase relationship ) due to the properties of NiTe$_2$ Dirac semimetal.

\begin{figure}
\includegraphics[width=\columnwidth]{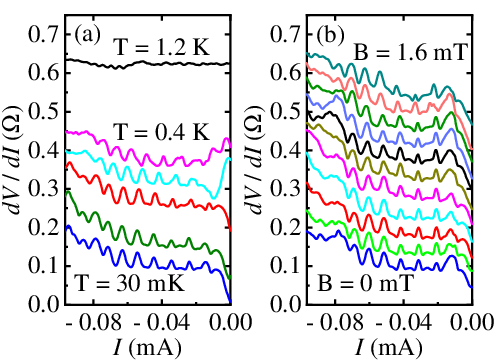}
\caption{(Color online) Integer and half-integer $dV/dI$ peaks for different temperatures (30~mK, 0.1~K, 0.2~K, 0.3~K, 0.4~K and 1.2~K)  (a) and magnetic fields (from 0 to 1.6~mT with .16~mT step) (b).  Shapiro steps are  not  sensitive to low $B<<B_c\approx 13$~mT magnetic fields in (b), while they are  moving to lower currents above 200~mK and disappear above 0.4~K in  (a), the half-integer steps disappear earlier. This behavior principally reflects the Josephson supercurrent behavior, as it can be expected for Shapiro steps.  Magnetic field is normal to the sample plane in  (b).  }
\label{fig4}
\end{figure}

The step positions are  stable at low temperatures in  Fig.~\ref{fig4} (a), while they are  moving to lower currents above 200~mK and disappear above 0.4~K.  $dV/dI$ peaks are also not  sensitive to low $B<<B_c\approx 13$~mT magnetic fields, see Fig.~\ref{fig4} (b).  This behavior principally reflects the critical current $I_c$ behavior in Fig.~\ref{fig2}, as it can be expected for Shapiro steps.  The half-integer steps disappear earlier in  Fig.~\ref{fig4} (a).

\section{Discussion}

First of all, there is no bulk superconductivity~\cite{Feng2021} for  NiTe$_2$ single crystal flakes at ambient pressure even at millikelvin temperatures. Bulk superconductivity  is known for pressurized Te-deficient NiTe$_2$~\cite{Feng2021}, but it can be ruled out in our case, since (i) bulk superconductivity is not observed for our NiTe$_2$ crystals according to four-point resistance data in Ref~\onlinecite{aunite}; (ii) X-ray spectroscopy reveals almost stoichiometric Ni$_{1-x}$Te$_2$  crystal with a slight Ni deficiency ($x$~$<$ 0.06); (iii) there is no external pressure in the present experiment. 

On the other hand, topological surface states carry the Josephson current on the pristine NiTe$_2$ surface~\cite{JDE,aunite,NiTe2_hinge}. Because of topological protection,  surface states  can  efficiently transfer the Josephson current, which  is reflected  in slow $I_c(T)$ decay in the inset to Fig.~\ref{fig2} (b). In this case, it is quite natural to observe integer Shapiro steps at $V=n hf/2e$,  as it should be expected for typical SNS junctions with trivial $2\pi$ periodicity in current-phase relationship $I_J\sim sin(\phi)$.

The specifics of our  In-NiTe$_2$-In junctions is the fact, that we do not  observe $4\pi$ periodicity in a.c. Josephson effect: the integer  $n=1$ Shapiro step is as strong as the $n=2$  in Fig.~\ref{fig3}, while the  maximum rf power value covers the range of $n=1$ disappearance in Refs.~\onlinecite{rfpan,rfli}. Moreover, the even sequence of Shapiro steps is usually observed~\cite{klapwijk_4pi} when frequency is below 1~GHz, similarly to our experiment.  $4\pi$ periodicity is connected with the helical nature~\cite{armitage} of topological surface states in Dirac semimetals~\cite{rfpan,rfli} and topological insulators~\cite{klapwijk_4pi}. Thus, we should explain both the presence of $\pi$ periodicity and why we do not see $4\pi$ one in comparison with~\cite{rfpan,rfli}. 

As the main experimental finding, fractional a.c. Josephson effect with $\pi$ periodicity of the current-phase relationship appears as half-integer $n=1/2,3/2,5/2$ and 7/2 Shapiro steps. For standard SNS junctions, higher weak-link transparency results in a more
skewed current-phase relationship~\cite{rfsnyder,CPP1,CPP2}, which is usually the origin of half-integer Shapiro steps. For our long $\xi<< L$ junctions, one could expect that it is high electron mobility in topological surface states that leads to skewed current-phase relationship~\cite{CPP2}.

Even in this case, one should expect inteference between one-dimensional channels~\cite{Vanneste,bezryadin,iran,frolov,chen}, which, for example, appears for  nontrivial Josephson current distribution in topological materials~\cite{inwte2,rfpan,rfsnyder}. 
Recently, $\pi$-periodic $sin(2\phi)$ current-phase relationship has been predicted and experimentally demonstrated for different realizations of superconducting quantum interference devices~\cite{sigeSQUID,grapheneSQUID,rhombusSQUID}. For short Cd$_3$As$_2$ Dirac semimetal junctions~\cite{rfpan,rfli}, interference ($\pi$ periodicity) could appear if both surface and bulk carriers transferred Josephson current in parallel. 

Each conduction channel, including the bulk Dirac fermions, Fermi-arc surface states, and topological hinge states, can be distinguished based on their different superconducting coherence lengths by increasing the channel length of the Josephson junction.   In our case, one cannot expect~\cite{kulik-long,dubos} the bulk Josephson current for 5~$\mu$m long In-NiTe$_2$-In  junctions, because of small bulk coherence length $\xi<<L$.  Helical surface states can transfer Josephson current for 1~$\mu$m distance~\cite{aunite}, while it is much smaller than $L=5$~$\mu$m for our In-NiTe$_2$-In  junctions. Also, we do not observe  $4\pi$ periodicity, which is the fingerprint of helical surface states. 

On the other hand, hinge states~\cite{hinge_theory,hingeCdAs_theory} have been theoretically predicted~\cite{NiTe2_hinge} and experimentally demonstrated~\cite{NiTe2_hinge}  in NiTe$_2$ type-II  Dirac semimetal. Interference of the hinge modes can explain all our observations: Josephson current for unprecidingly long In-NiTe$_2$-In junctions, and observation of $\pi$ periodicity without the $4\pi$ one. (i)  The coherence length should be longer in the one-dimensional hinge channel than that in the surface and bulk states, giving rise to the supercurrent being dominated by the hinge channels in long Josephson junctions~\cite{hingeCdAs_experiment}. (ii) For one-dimensional hinge states, one can expect  interference of supercurrent through the states localized at the intersections between surfaces, i.e. fractional a.c. Josephson effect with $\pi$ periodicity~\cite{Vanneste,bezryadin,iran,frolov,chen}. (iii) Topological hinge states  inherit the chiral property of the Chern insulator edge states~\cite{hinge_review,hinge_theory}, so they should not demonstrate $4\pi$ periodicity of the current-phase relationship.

Thus, our experiment can be considered not only as demonstration of transport through the topological  hinge states in  NiTe$_2$ Dirac semimetal, but also as the confirmation of their chiral property.

\section{Conclusion}

As a conclusion,   we experimentally investigate Josephson current between two 5~$\mu$m spaced superconducting indium leads, coupled to a NiTe$_2$ single crystal flake, which is  a type-II Dirac semimetal.  Under microwave irradiation, we demonstrate a.c. Josephson effect at millikelvin temperatures as a number of Shapiro steps. In addition to the integer ($n=1,2,3,4...$) steps, we observe fractional ones at half-integer values $n=1/2,3/2,5/2$ and 7/2, which corresponds to $\pi$ periodicity of current-phase relationship. In contrast to previous investigations, we do not  observe $4\pi$ periodicity (disappearance of the odd $n=1,3,5...$ Shapiro steps), while the latter is usually considered as a fingerprint of helical surface states in Dirac semimetals and topological insulators.  We argue, that our experiment confirms Josephson current through the topological hinge states in  NiTe$_2$: since one can exclude bulk supercurrent in 5~$\mu$m long Josephson junctions, interference of the hinge modes is responsible for the $\pi$ periodicity, while stable odd Shapiro steps reflect chiral character of the topological  hinge states.    

\acknowledgments
We wish to thank  S.S.~Khasanov for X-ray sample characterization, and  A.N. Nekrasov for the energy-dispersive X-ray spectroscopy. We gratefully acknowledge financial support by the RF State task.

\end{document}